\documentclass[11pt]{article}
\usepackage{amssymb,amsmath,cite,geometry}
\usepackage[usenames,dvipsnames]{color}
\catcode`\@=11

\@addtoreset{equation}{section}
\def\theequation{\arabic{section}.\arabic{equation}}
     
\catcode`\@=11
\def\thesection{\arabic{section}.}

\def\appendix{\setcounter{section}{0}
        \def\thesection{Appendix.}
        \def\theequation{\Alph{section}.\arabic{equation}}}
\def\section{\@startsection{section}{1}{\z@}{3.5ex plus 1ex minus
   .2ex}{2.3ex plus .2ex}{\large\bf}}
     
\long\def\@makefntext#1{\parindent 0cm\noindent
\hbox to 1em{\hss$^{\@thefnmark}$}#1}

\newcommand{\captionfonts}{\small}
\makeatletter  
\long\def\@makecaption#1#2{%
  \vskip\abovecaptionskip
  \sbox\@tempboxa{{\captionfonts #1: #2}}%
  \ifdim \wd\@tempboxa >\hsize
    {\captionfonts #1: #2\par}
  \else
    \hbox to\hsize{\hfil\box\@tempboxa\hfil}%
  \fi
  \vskip\belowcaptionskip}
\makeatother   

\begin{document}
\begin{titlepage}
\vspace{.5in}
\begin{flushright}
June 2014\\  
\end{flushright}
\vspace{.5in}
\begin{center}
{\Large\bf
 Lorentz Invariance in Shape Dynamics}\\  
\vspace{.4in}
{S.~C{\sc arlip}\footnote{\it email: carlip@physics.ucdavis.edu}\\
 H{\sc enrique} G{\sc omes}\footnote{\it email: gomes.ha@gmail.com}\\
       {\small\it Department of Physics}\\
       {\small\it University of California}\\
       {\small\it Davis, CA 95616}\\{\small\it USA}}
\end{center}

\vspace{.5in}
\begin{center}
{\large\bf Abstract}
\end{center}
\begin{center}
\begin{minipage}{4.75in}
{\small 
Shape dynamics is a reframing of canonical general relativity in which
time reparametrization invariance is ``traded'' for a local conformal
invariance.  We explore the emergence of Lorentz invariance in this
model in three contexts: as a maximal symmetry, an asymptotic
symmetry, and a local invariance.
}
\end{minipage}
\end{center}
\end{titlepage}
\addtocounter{footnote}{-2}

It is a central premise of modern physics that we live in a four-dimensional 
spacetime.  Space and time, although not identical, are intertwined: there is 
no preferred synchronization of spatially separated clocks, and a fundamental 
symmetry, diffeomorphism invariance, mixes space and time.  The greatest 
triumph of this worldview is general relativity, which elegantly describes 
gravity as an effect of the curvature of spacetime.

It is hard to imagine the discovery of general relativity without the benefit 
of such a four-dimensional picture.  In hindsight, though, there is an 
alternative description.  In the Hamiltonian formulation \cite{ADM}, 
the dynamical variables of general relativity are a spatial metric $g_{ij}$ 
and its conjugate momentum $\pi^{ij}$, which, like the variables in any 
canonical theory, trace out a curve in phase space.  The time-time and 
mixed space-time components of the metric appear only as Lagrange 
multipliers.  Spatial diffeomorphism invariance remains, but the time 
refoliations are replaced by ``surface deformations'' acting on $g_{ij}$
and $\pi^{ij}$, which are equivalent to diffeomorphisms only on shell
\cite{Teitelboim}.  The
spacetime picture emerges almost miraculously, as a consequence of the
structure of the constraint equations.

Such miracles should not be taken lightly.  But one can still ask whether
the canonical theory can be modified in an interesting way.  It can.
Shape dynamics \cite{shape1,shape2,shape3} is a theory of gravity in 
which the surface deformations of standard canonical general relativity 
are ``traded'' for local scale transformations.  In many circumstances%
---specifically, for solutions that can be described as spacetimes that 
admit a global constant mean curvature time-slicing---shape dynamics 
is equivalent to general relativity.  But  there are circumstances in which 
the theories do not agree, and the different constraint structures suggest 
very different approaches to quantization.

To treat shape dynamics as a fundamental theory, though, we must confront 
a rather glaring problem.  While general relativity may be the greatest
triumph of the four-dimensional picture of the Universe, spacetime is also
the natural setting for special relativity.  How can Lorentz invariance arise
in a three-dimensional theory like shape dynamics?  One answer is trivial,
but uninformative: since shape dynamics is (often) equivalent to general 
relativity, it must have the same symmetries.  Surely, though, there must be
a more physical explanation.

In this paper, we examine the emergence of Lorentz invariance in shape
dynamics in three settings: as a maximal symmetry of solutions, as an
asymptotic symmetry of asymptotically flat solutions, and as a local
invariance.  We show how a combination of kinematics and dynamics
leads to Lorentz invariance, and demonstrate both the strength of the
symmetry and some of the possible ways to break it.

\section{Shape dynamics}

The phase space of canonical general relativity is parametrized by fields
$({\bar g}_{ij},{\bar\pi}^{ij})$ satisfying two sets of constraints: the 
momentum (or ``diffeomorphism'') constraints%
\footnote{We shall work in canonical formalism throughout this paper; 
unless otherwise stated, all geometrical objects (such as the scalar curvature 
$\bar R$) will be spatial tensors defined in terms of the three-metric on a slice 
$\Sigma$.}
\begin{align}
{\bar{\cal H}}[\xi] = -\int d^3x\,({\bar\nabla}_i{\bar\xi}_j + {\bar\nabla}_j{\bar\xi}_i)
   {\bar\pi}^{ij}
\label{a1}
\end{align}
which generate spatial diffeomorphisms along the vector field 
$\xi^i = {\bar g}^{ij}{\bar\xi}_j$,  and the scalar (or ``Hamiltonian'') constraint,
\begin{align}
{\bar{\cal S}}[\xi^\perp] = \int d^3x\,\xi^\perp\left[\frac{1}{\sqrt{{\bar g}}}
    \left({\bar\pi}^{ij}{\bar\pi}_{ij} - {\bar\pi}^2\right)
     - \sqrt{{\bar g}}{\bar R}\right]
\label{a2}
\end{align}
which generates surface deformations parametrized by the function 
$\xi^\perp$.  The algebra of constraints closes under Poisson brackets,
provided the parameters $(\xi^\perp,\xi^i)$ obey the surface deformation 
algebra
\begin{align}
&\{\xi,\eta\}_{\scriptscriptstyle\mathit{Surf}}^\perp 
      = \xi^i{\bar\nabla}_i\eta^\perp -\eta^i{\bar\nabla}_i\xi^\perp \nonumber\\
&\{\xi,\eta\}_{\scriptscriptstyle\mathit{Surf}}^i
      = \xi^k{\bar\nabla}_k\eta^i - \eta^k{\bar\nabla}_k\xi^i
      + {\bar g}^{ik}(\xi^\perp{\bar\nabla}_k\eta^\perp - \eta^\perp{\bar\nabla}_k\xi^\perp ) 
\label{a2a}
\end{align}
On shell, the corresponding transformations are equivalent to spacetime
diffeomorphisms generated by a vector field ${\tilde\xi}^\mu$ with 
$\displaystyle(\xi^\perp,\xi^i) = (N{\tilde\xi}^t, {\tilde\xi}^i + N^i{\tilde\xi}^t)$.
 
To obtain shape dynamics\footnote{We consider here 
only that case of a spatially open manifold.  The spatially compact case 
is technically more difficult, since one must treat total-volume-preserving
 rescalings separately; see \cite{shape1} for details.}
we first use the ``Stueckelberg trick'' \cite{Stuec} to make the theory invariant
under local spatial rescalings: we introduce a scalar field $\phi$ and write
\begin{align}
{\bar g}_{ij} = e^{4\phi}g_{ij}, \quad {\bar\pi}^{ij} = e^{-4\phi}\pi^{ij}
\label{a3}
\end{align}
under which the constraints become
\begin{subequations}
\begin{align}
{\cal H}[\xi] &= -\int d^3x\,\left[(\nabla_i\xi_j + \nabla_j\xi_i) \pi^{ij}
      + 4\xi^k\partial_k\phi\,\pi\right]
\label{a4a}\\
{\cal S}[\xi^\perp] &=  \int d^3x\,\xi^\perp\left[\frac{e^{-6\phi}}{\sqrt{g}}
    \left(\pi^{ij}\pi_{ij} -  \pi^2\right)
     - e^{2\phi}\sqrt{g}\left(R - 8\Delta\phi -8 \nabla_i\phi\nabla^i\phi\right)\right]
\label{a4b}
\end{align}
\end{subequations}
where $\Delta$ is the Laplacian $\nabla^i\nabla_i$.
The theory is then trivially invariant under local rescalings of $(g_{ij},\pi^{ij})$
accompanied by shifts of $\phi$, a symmetry generated by the new constraint
\begin{align}
{\cal Q}[\rho] = \int d^3x \rho(\pi_\phi - 4 \pi) ,
\label{a5}
\end{align}
where $\pi_\phi$ is the momentum conjugate to $\phi$ and, as usual, 
$\pi = g_{ij}\pi^{ij}$.  
This extended ``linking theory'' \cite{shape2} is, of course, equivalent to general relativity,
which can be recovered by choosing the gauge $\phi=0$.  Technically, this
gauge condition does not commute with the constraint $\cal Q$, so we must
also solve the equation ${\cal Q}=0$; but this merely fixes $\pi_\phi$, which
appears nowhere else.

For shape dynamics, we instead choose the gauge $\pi_\phi=0$.  By
(\ref{a5}), this is a choice of maximal slicing, implying that the trace of
the extrinsic curvature vanishes.  This gauge condition commutes with $\cal Q$, 
so local rescalings remain as a symmetry.  It does not, however, commute with 
the scalar constrain $\cal S$, which must therefore be solved.  

More precisely, the Poisson bracket $\{{\cal S}[N],\pi_\phi\}$ vanishes only 
when $N$ obeys the ``lapse-fixing equation''
\begin{align}
\Delta N + 2\nabla^i\phi\nabla_iN 
    -\frac{e^{-8\phi}}{g}\left(\pi^{ij}\pi_{ij} - \frac{\pi^2}{2} \right)N = 0
\label{a6}
\end{align}
A solution $N_0$ of this equation may be said to propagate the gauge choice,
and ${\cal S}[N_0]$ remains a first class constraint.  As we shall see below,
$N_0$ is typically unique if $\pi^{ij}\ne0$, but may not be unique if
$\pi^{ij}$ vanishes.  

The vanishing of the remaining scalar constraint gives an equation for 
$\Omega=e^\phi$,
\begin{align}
\Delta\Omega - \frac{1}{8}R\Omega 
    + \frac{1}{8}\frac{1}{g}\pi^{ij}\pi_{ij}\Omega^{-7}=0 
\label{a7}
\end{align}
which, given suitable boundary conditions, has a unique solution.  Note that
(\ref{a6}) and (\ref{a7}) both require boundary conditions in order to be
well-defined.  This is an early indication of one of the main differences between
shape dynamics and general relativity: shape dynamics is more ``global,''
and some questions that can be answered purely locally in general relativity
require global information in shape dynamics.  Much of this difference can
be traced back to the fact that shape dynamics is formulated from the start
as a canonical theory, for which global information is needed to define the
phase space.

We now have a theory with a phase space parametrized by $(g_{ij},\pi^{ij})$, 
with constraints ${\cal H}[\xi]$ and $ {\cal Q}[\rho]$ given by (\ref{a4a})
and (\ref{a5}), and a Hamiltonian ${\cal S}[N_0]$ determined by (\ref{a4b}). 
We emphasize that ${\cal S}[N_0]$ is a Hamiltonian only when $N_0$ solves 
the lapse-fixing equation (\ref{a6}); otherwise, it does not commute with
the gauge condition $\pi_\phi = 0$, and should be eliminated by solving 
(\ref{a7}) for $\phi$.  For a spatially open manifold, additional boundary
terms are needed to make the Poisson brackets of these constraints
well-defined; we discuss these briefly in section \ref{asymp}  When 
$\phi$ (determined by (\ref{a7})) and $N$ (determined by (\ref{a6})) 
have positive unique solutions, the shape dynamics solutions are 
equivalent to those of general relativity in the maximal slicing gauge, 
with a metric (\ref{a3}).

With this background, we can now turn to our primary problem:
understanding the emergence of Lorentz invariance in this setting.

\section{Maximal symmetries}

In general relativity, symmetries are generated by Killing vectors, which are
usually described in the full (3+1)-dimensional setting.  Shape dynamics, on
the other hand, is a canonical theory.  It is therefore helpful to start by looking
at the canonical version of the Killing equation in general relativity.  

Given a lapse and shift ${\cal N}^\mu = (N,N^i)$ and a Killing vector 
${\tilde\chi}^\mu$, define Killing surface deformation parameters 
$(\chi^\perp,\chi^i) = (N{\tilde\chi}^t, {\tilde\chi}^i + N^i{\tilde\chi}^t)$.
Then it is straightforward to show that the Killing equation takes the
form \cite{KID,Carlipbh}
\begin{subequations}
\begin{align}
\delta_\chi g_{ij} &= \nabla_i\chi_j + \nabla_j\chi_i 
      + \frac{2}{\sqrt{g}}\chi^\perp\left(\pi_{ij} - \frac{1}{2}g_{ij}\pi\right) = 0
      \label{b1a}\\
\delta_\chi \pi^{ij} &= \chi^\perp\sqrt{g}\left(R^{ij}-\frac{1}{2}g^{ij}R\right)
      +\frac{{\chi^\perp}}{\sqrt g}\left(2(\pi^{ik}\pi^j{}_k-\pi\pi^{ij})
      -\frac{1}{2}g^{ij}(\pi^{k\ell}\pi_{k\ell}-\frac{1}{2}\pi^2)\right) \nonumber\\
      &\quad- \sqrt {g}\left(\nabla^i\nabla^j\chi^\perp - g^{ij}\Delta\xi^\perp\right)
      -\nabla_k(\chi^k\pi^{ij}) + \pi^{ik}\nabla_k\chi^j  + \pi^{jk}\nabla_k\chi^i= 0
      \label{b1b}\\
\partial_t\chi^\mu &= -\{{\cal N},\chi\}_{\scriptscriptstyle\mathit{Surf}}^\mu
\label{b1c}
\end{align}
\end{subequations}
where the brackets in (\ref{b1c}) are the surface deformation brackets (\ref{a2a}).
The first two equations define Killing initial data; the surface deformation
brackets then propagate this initial data to the future.  The Killing equations are 
diffeomorphism invariant, so the system of equations (\ref{b1a}--\ref{b1c}) will 
hold for any time-slicing and any choice of lapse and shift.

In particular, if we start with flat initial data $g_{ij}=\delta_{ij},\ \pi^{ij}=0$,
then (\ref{b1a}--\ref{b1b}) tells us that at $t=0$, $\chi^i$ is a three-dimensional 
Killing vector and $\partial_i\partial_j\chi^\perp = 0$, i.e.,
\begin{align}
&\chi^\perp(0) = a^0 + \omega^0{}_jx^j \nonumber\\
&\chi^i(0) = a^i + \omega^i{}_jx^j 
\label{b2}
\end{align}
where $\omega_{ij}$ is an antisymmetric constant.  If we choose a lapse and shift
$N=1$, $N^i=0$, the propagation equation (\ref{b1c}) then reduces to
$\partial_t\chi^\perp = 0,\ \partial_t\chi^i = \omega^0{}_i$, and thus
\begin{align}
\chi^\mu(t) = a^\mu + \omega^\mu{}_\nu x^\nu
\label{b3}
\end{align}
giving the expected Poincar{\'e} symmetry.  It is easy to check that although the
surface deformation brackets are not the usual commutators, these vectors obey 
the standard Poincar{\'e} algebra.  This is, in fact, a general result: it follows from
Appendix A of \cite{Brown} that for Killing vectors, the surface deformation
brackets are equal to the usual commutators.

Let us now apply the same arguments to shape dynamics.  We should first clarify
one slightly subtle point.  We are interested in symmetries of solutions, not
just of initial data.  In particular, consider data at time $t_0$ that generate a 
particular solution.  If a transformation takes us from this data to data at a 
different time $t_1$ that generate the \emph{same} solution, this is a symmetry.
For general relativity, this is not an issue, since the Hamiltonian is itself a
symmetry generator.  For shape dynamics, on the other hand, ${\cal S}[N_0]$
is often treated as a genuine Hamiltonian, which must also be incorporated 
in the analysis.

We thus begin with a generator of the form
\begin{align}
H[\xi^\perp,\xi^i,\xi_s] = {\cal S}[\xi^\perp] + {\cal H}[\xi^i] + {\cal Q}[\xi^s]
\label{b4}
\end{align}
where $\xi^\perp$ is restricted to obey the lapse-fixing equation (\ref{a6}).
We first need the analog of the surface deformation brackets, which can be
obtained by computing the Poisson bracket $\{H[\xi],H[\eta]\}$ and writing
the result as $H[\{\xi,\eta\}_{\scriptscriptstyle\mathit{Shape}}]$.  A long
but unexciting calculation yields  
\begin{align}
&\{\xi,\eta\}_{\scriptscriptstyle\mathit{Shape}}^\perp 
      \circeq \xi^i\nabla_i\eta^\perp -\eta^i\nabla_i\xi^\perp 
      - 2\xi^s\eta^\perp + 2\eta^s\xi^\perp\nonumber\\
&\{\xi,\eta\}_{\scriptscriptstyle\mathit{Shape}}^i
      = \xi^k\nabla_k\eta^i - \eta^k\nabla_k\xi^i
      + g^{ik}(\xi^\perp\nabla_k\eta^\perp - \eta^\perp \nabla_k\xi^\perp ) 
\label{b5}\\
&\{\xi,\eta\}_{\scriptscriptstyle\mathit{Shape}}^s 
      = \xi^i\nabla_i\eta^s -\eta^i\nabla_i\xi^s \nonumber
\end{align}
There is once again an important subtlety, however.  As we have stressed,
${\cal S}[N_0]$ is a Hamiltonian only if $N_0$ satisfies the lapse-fixing
equation ({\ref{a6}).  In general, the right-hand side of the first equation
in (\ref{b5}) will not obey this condition.  We use the symbol $\circeq$ to
mean ``equal when projected onto the space of solutions of the
lapse-fixing equation''; that is, the left-hand side of the first equation
in (\ref{b5}) is equal to the unique solution of (\ref{a6}) that has the same
boundary values as the right-hand side.  

Next, as in (\ref{b1a}--\ref{b1b}). we need to set the variation of the
initial data to zero.  We find 
\begin{subequations}
\begin{align}
\delta_\chi g_{ij} &= \nabla_i\chi_j + \nabla_j\chi_i 
      + \frac{2}{\sqrt{g}}e^{-6\phi_0}\chi^\perp\left(\pi_{ij} - \frac{1}{2}g_{ij}\pi\right) 
       + 4\chi^s g_{ij}= 0
      \label{b6a}\\
\delta_\chi \pi^{ij} &= \chi^\perp e^{2\phi_0}\sqrt{g}\left((R^{ij}-\frac{1}{2}g^{ij}R)
      - 2\nabla^i\nabla^j\phi_0 + 4\nabla^i\phi_0\nabla^j\phi_0 +2g^{ij}\Delta\phi_0\right)\nonumber\\
      &\quad+\frac{1}{\sqrt g}e^{-6\phi_0}\chi^\perp\left(2(\pi^{ik}\pi^j{}_k-\pi\pi^{ij})
      -\frac{1}{2}g^{ij}(\pi^{k\ell}\pi_{k\ell}-\frac{1}{2}\pi^2)\right)\nonumber\\
      &\quad- \sqrt {g}\biggl(\nabla^i\nabla^j\chi^\perp - g^{ij}\Delta\chi^\perp
      - 2\nabla^i\phi_0\nabla^j\chi^\perp - 2\nabla^j\phi_0\nabla^i\chi^\perp\biggr)\nonumber\\
      &\quad-\nabla_k(\chi^k\pi^{ij}) + \pi^{ik}\nabla_k\chi^j  + \pi^{jk}\nabla_k\chi^i -4\chi^s\pi^{ij}= 0
      \label{b6b}
\intertext{where $\phi_0$ is again determined by (\ref{a7}).  To propagate the  
parameters $\chi$,  we should replace the lapse and shift by a triple 
${\widetilde{\cal N}} = (N,N^i,\rho)$, where $\rho$ is a Lagrange multiplier 
for the conformal constraint ${\cal Q}$; then}
\partial_t\chi &= -\{{\widetilde{\cal N}},\chi\}_{\scriptscriptstyle\mathit{Shape}} 
\label{b6c}
\end{align}
\end{subequations}
 
Let us again consider flat initial data, $g_{ij}=\delta_{ij},\ \pi^{ij}=0$.  A suitable
solution of (\ref{a7}) is $\phi_0=0$, and (\ref{b6b}) again tells us that 
$\partial_i\partial_j\chi^\perp = 0$.  Now, however, (\ref{b6a}) implies 
that $\chi^i$ can be a three-dimensional \emph{conformal} Killing vector, i.e.,%
\begin{align}
&\chi^\perp(0) = a^0 + \omega^0{}_jx^j\nonumber\\
&\chi^i(0) = a^i + \omega^i{}_jx^j + kx^i + (c^i(x_kx^k) - 2(c_kx^k) x^i) \label{b7}\\
&\chi^s(0) = -\frac{k}{2} + c_kx^k \nonumber
\end{align}
The propagation equation (\ref{b6c}) is then easily integrated, yielding
\begin{align}
&\chi^t(t) = a^0 + \omega^0{}_jx^j + kt -2(c_kx^k) t\nonumber\\
&\chi^i(t) = a^i + \omega^i{}_jx^j 
    + \omega^i{}_0t + kx^i + (c^ix^2 - 2(c_kx^k)x^i) \label{b8}\\
&\chi^s(t) = -\frac{k}{2} + c_kx^k \nonumber
\end{align}

This is \emph{almost} the full four-dimensional conformal group.  One 
transformation is missing, though---the time component of 
the special conformal transformations, $\chi^\mu = (c(x^2+t^2),2ctx^i)$,
is absent.  This four-dimensional conformal Killing vector can be 
generated by the propagation equation (\ref{b6c}), but the corresponding 
initial data, $\chi^\perp = cx^2$, fails to preserve the momentum 
$\pi^{ij}$, and does not satisfy the lapse-fixing equation.  The 
corresponding transformation does not take maximal slices to maximal 
slices, and is thus not permitted in shape dynamics.

Just as the surface deformation brackets were equivalent to ordinary
commutators for Killing vectors, the extra $\xi^s$ term in  
(\ref{b5}) makes the shape dynamics brackets equivalent to  ordinary 
commutators for conformal Killing vectors \cite{Brown}.  But because 
of the missing special conformal generator, the transformations 
generated by (\ref{b8}) fail to close.  As first pointed out in \cite{asymp}, 
the symmetries are inconsistent, and one must restrict to a subgroup.  

At first sight, this is a rather peculiar situation.  Its source may be 
traced back to the global nature of shape dynamics.  For a spatially 
open manifold, boundary conditions at infinity are required to define 
a sensible phase space.  The usual asymptotically flat boundary 
conditions forbid even the spatial special conformal transformations,
a restriction that is invisible if one only looks at the local behavior. 
When one restricts to symmetries that preserve boundary conditions,
one obtains the Poincar{\'e} group alone, just as in general relativity.
Shape dynamics thus reproduces Lorentz invariance as a maximal 
symmetry, but we now require global information to obtain this result.

It is an interesting open question whether boundary conditions for shape
dynamics could be enlarged to allow a larger group of symmetries.
We have not yet found a way to do so, but we cannot rule out the
possibility.  If such boundary conditions exist, they could provide a
new arena in which shape dynamics differs from general relativity.

\section{Asymptotic Lorentz invariance \label{asymp}}

A second place in which Lorentz invariance appears in general relativity 
is as an asymptotic symmetry of asymptotically flat metrics.  In shape
dynamics, this aspect has been discussed in detail in \cite{asymp},
so we will merely summarize the results here.

An analysis of asymptotic symmetries requires two ingredients: a set
of boundary conditions or fall-off conditions for the fields, and a set of 
boundary terms that make the symmetry generators ``differentiable''%
---that is, that make their variations well-defined even after integration 
by parts \cite{Regge}.  The two are not independent: the boundary terms 
are largely fixed by the boundary conditions, which in turn must be 
chosen to allow boundary terms to be defined.  The boundary terms then
give ``charges,'' most famously the ADM mass, associated with the
corresponding symmetries.

Following \cite{Regge,Beig}, we chose boundary conditions for the 
spatial metric and momentum of the form
\begin{subequations}
\begin{alignat}{3}
&g_{ij} \rightarrow \delta_{ij} + \mathcal{O}(r^{-1})^{\prime\prime} \qquad
&& \pi^{ij} \rightarrow \mathcal{O}(r^{-2})' \label{c1}\\
\intertext{while allowing looser restrictions on the Lagrange multipliers,} 
&N \rightarrow \alpha_{(a)}x^{(a)} + c + \mathcal{O}(1)'\qquad\qquad
&&N^i \rightarrow \beta^{(a)}\delta^i_{(a)} + \mu_{(a)(b)}x^{[(a)}\delta^{(b)]i }
    + \mathcal{O}(1)'\nonumber\\
&\rho \rightarrow \mathcal{O}(r^{-1})^{\prime\prime}
\label{c2}
\end{alignat}
\end{subequations}
Here, the notation $\mathcal{O}(r^n)^{\prime\prime}$ denotes a function
that falls off as an even parity term of order $r^n$ plus arbitrary terms 
of order $r^{n-1}$; $\mathcal{O}(r^n)^{\prime}$ denotes a function that
falls off as an odd parity term of order $r^n$ plus arbitrary terms of order 
$r^{n-1}$.  Indices in parentheses in (\ref{c2}) may be viewed as basis indices, 
labeling independent fall-off conditions.  These boundary  conditions allow 
solutions of the lapse-fixing equation (\ref{a6}), and the asymptotic form of 
the Lagrange multipliers is chosen to match the symmetries of Minkowski 
space found in the preceding section.

We must next determine an appropriate boundary term $\mathcal{B}[N,N^i,\rho]$
to add to the generator $H[N,N^i,\rho]$ of (\ref{b4}) to make its variation 
well-defined.  As Regge and Teitelboim first pointed out \cite{Regge}, the
bulk term by itself is not sufficient: its variation requires integration by
parts, leading to boundary integrals that must be cancelled in order for
functional derivatives to exist.  An added subtlety is that the boundary term
$\mathcal{B}[N,N^i,\rho]$ must itself be finite; this can restrict the allowable 
boundary conditions.

The variation is evaluated in \cite{asymp}, where it is shown that for the
boundary conditions (\ref{c1})--(\ref{c2}),
\begin{align}
\mathcal{B}[N,N^i,\rho] &= \int_{\partial\Sigma}r^i\left( 8N\partial_i\Omega
   + 2\xi_j\pi^j{}_i + \sqrt{h}(N\partial_\ell g_{jk} -\partial_\ell Ng_{jk})
   (\delta^\ell_ig^{jk} - \delta^j_ig^{\ell k})\right)d^2y
\label{c3}
\end{align}
where $r^i$ is a unit normal at the boundary, $\sqrt{h}$ is the induced 
volume element, and $\Omega = e^\phi$ is the solution of the constraint 
(\ref{a7}).   All but the first term in (\ref{c3}) are the standard ADM boundary 
terms.  The first term is new, however, and has the effect of making the 
ADM mass Weyl invariant.  That is, for transformations
\begin{align}
(g_{ij},\pi^{ij})\rightarrow(e^{4\psi}g_{ij},e^{-4\psi}\pi^{ij})
\label{c4}
\end{align}
(with $\psi \rightarrow0$ at infinity to preserve the boundary conditions),
the standard ADM mass will acquire new terms involving radial derivatives
of $\psi$.  These are exactly canceled by the first term in the shape dynamics
Hamiltonian, making the mass invariant.

It remains for us to determine the algebra of the generators 
$(H+\mathcal{B})[N,N^i,\rho]$.  On general grounds \cite{BrownHenneaux}, 
one expects the Poisson brackets to be those of the bulk---and thus
given by the ``shape dynamics'' brackets (\ref{b5})---up to a possible
central term.  It is shown in \cite{asymp} that no central term occurs.
One can then read off the algebra from the asymptotic form (\ref{c2}),
and confirm that it is precisely the Poincar{\'e} algebra, with
translations parametrized by $(c,\beta^{(a)})$, rotations parametrized
by $\mu_{(a)(b)}$, and boosts parametrized by $\alpha_{(a)}$.

As in the preceding section, we can ask whether it is possible to enlarge
our boundary conditions to allow conformal transformations as
asymptotic symmetries.  While we again cannot completely rule out 
this possibility, it seems very difficult: such boundary conditions
typically require that $\Omega$ grow rapidly at infinity, and for such
growth there seems to be no way to define finite boundary terms for
the generator of the symmetry. A further hurdle is that conformal 
factors that grow at infinity may not respect the naive radial power 
expansion of the metric fields,  in which case they would not preserve 
the boundary conditions.   

\section{Local Lorentz invariance}

The final place that Lorentz invariance appears in general relativity is as
``local Lorentz invariance.''  This term has several interpretations, 
though, and one must be a bit careful about what one means.
In its simplest form, local Lorentz invariance is the statement that one
can define an orthonormal tetrad in a neighborhood of any point, and 
that different choices of tetrads are related Lorentz transformations.
But this is really a statement of mathematics rather than physics: it is
trivially true for a manifold with a Lorentzian metric.  Since 
shape dynamics gives a prescription for constructing such a manifold%
---at least in regions where the lapse determined by (\ref{a6}) does not
vanish---this adds little physical content.\footnote{It is important, of
course, to understand the places where this procedure breaks down.
This occurs, for instance, at the horizon of a stationary black hole
\cite{SDbh,SDbh2}, where the differences between general relativity
and shape dynamics may be partially understood as a breakdown of 
local Lorentz invariance.}  A somewhat stronger version is a demand that
couplings to matter respect this invariance.  But the tetrad is needed only
to incorporate fermions.  The coupling of fermions in shape dynamics
has not been fully worked out (see \cite{Soo} for some related ideas), 
but presumably we should not be satisfied with a definition of local
Lorentz invariance that applies only to fermions.

Physically, what we most commonly mean by local Lorentz invariance is  
closely tied to the equivalence principle \cite{Will}.  It is the statement 
that in a sufficiently small region of spacetime, we can choose a freely 
falling reference frame in which the metric is locally indistinguishable 
from the Minkowski metric.  In other words, local Lorentz invariance
can be characterized by the existence of coordinates in which 
freely falling objects move along straight lines, and in which null 
geodesics determine a standard light cone structure.  The key question 
in shape dynamics, then, is whether such a choice is possible---now 
through a combination of three-dimensional coordinate transformations 
and local conformal transformations---or whether the existence of 
a preferred time-slicing prevents the required choices.

Ideally, we should start by deriving the equations of motion for a
freely falling object from first principles, for instance by looking 
at the geometric optics limit for matter fields coupled to shape dynamics.
Fortunately, though, most of this work is done for us.  We know that in
a ``typical'' region of space, the solutions of shape dynamics are
equivalent to those of general relativity in a particular gauge.
But we have also known since the work of Einstein, Infeld and Hoffmann
\cite{EIH} that in any such solution, test particles move along
geodesics.

We therefore start with the geodesic equation, which
can be written in (3+1)-dimensional form as \cite{Gourgoulhon}
\begin{multline}
\frac{d^2x^i}{dt^2} =
    - \frac{1}{N}\Bigl[\frac{1}{\sqrt{{\bar g}}}\,{\bar g}_{j\ell}{\bar g}_{km}{\bar\pi}^{\ell m}%
    \left(\frac{dx^j}{dt} + N^j\right) \left(\frac{dx^k}{dt} + N^k\right)
    - \left(2\frac{dx^j}{dt} + N^j\right)\partial_jN - \frac{\partial N}{\partial t}\Biggr]%
    \frac{dx^i}{dt} \\
    + 2\frac{N}{\sqrt{{\bar g}}}\left[{\bar g}_{j\ell}{\bar\pi}^{i\ell} 
    - \frac{N^i}{N}\frac{N^k}{N}{\bar g}_{k\ell}{\bar g}_{jm}{\bar\pi}^{\ell m}
    - \sqrt{{\bar g}}\,{\bar\nabla}_j\left(\frac{N^i}{N}\right)\right]\frac{dx^j}{dt}\\
    - \left({\bar\Gamma}^i_{jk} + \frac{1}{\sqrt{{\bar g}}}\frac{N^i}{N}
     {\bar g}_{j\ell}{\bar g}_{km}{\bar\pi}^{\ell m}\right)
    \frac{dx^j}{dt}\frac{dx^k}{dt} - N{\bar g}^{ij}\partial_jN 
    - \frac{\partial N^i}{\partial t} - N^j{\bar\nabla}_jN^i\\
    + 2N{\bar g}_{j\ell}{\bar\pi}^{i\ell}N^j
    - \frac{N^i}{N}\left(\frac{1}{\sqrt{{\bar g}}}\,{\bar g}_{j\ell}{\bar g}_{km}{\bar\pi}^{\ell m}N^jN^k
    - \frac{\partial N}{\partial t} - N^j\partial_jN\right)   \label{d1}
\end{multline}
where ${\bar g}_{ij}$ and ${\bar\pi}^{ij}$ are given by (\ref{a3}), and
we have used the fact that we are on a slice of vanishing mean curvature.
In this expression, the lapse $N$ is determined by the lapse-fixing equation,
but the shift $N^i$ is arbitrary.  We are also free to make an arbitrary spatial
coordinate transformation and an arbitrary local Weyl transformation.  Note,
though, that the parameters of these transformations should not depend on
the particle position or velocity, since we want the right-hand side of (\ref{d1})
to vanish for \emph{all} geodesics at some spacetime point $p$.

It is convenient to sort the terms in (\ref{d1}) by the power at which the velocity
appears.  The last few terms are velocity-independent, and we can make them 
vanish at $p$ by choosing
\begin{align}
N^i = 0, \qquad  
\frac{\partial N^i}{\partial t} + N^j{\bar\nabla}_jN^i = -N{\bar g}^{ij}\partial_jN
\qquad\hbox{at $p$}
\label{d2}
\end{align}
The second equality in this equation resembles a force law: $N{\bar g}^{ij}\partial_jN$  
has the form of a Newtonian gravitational force, and our requirement is that the 
shift vector follow this force.

We can remove further terms in (\ref{d1}) by choosing spatial coordinates 
such that
\begin{align}
{\bar\Gamma}^i_{jk} = 0 \qquad\hbox{at $p$}
\label{d3}
\end{align}
and requiring that
\begin{align}
\sqrt{{\bar g}}\,{\bar\nabla}_j\left(\frac{N^i}{N}\right) = {\bar g}_{j\ell}{\bar\pi}^{i\ell}  
   \qquad\hbox{at $p$}
\label{d4}
\end{align}
Note that on shell, (\ref{d4}) implies that $\partial_t {\bar g}_{ij}=0$ at $p$.

Neither of these expressions is integrable, of course; just as in general relativity,
these choices can only be imposed at a point.  As usual, the integrability
condition for (\ref{d3}) requires vanishing of the spatial curvature tensor,
while the integrability condition for (\ref{d4}) requires the vanishing of a term 
of the form $\nabla_iK_{jk} - \nabla_jK_{ik} \sim {}^{\scriptsize(4)}R^n{}_{kij}$.

The first term on the left-hand side of (\ref{d1}), however, cannot be
eliminated by shape dynamics transformations, and we thus have
\begin{align}
\frac{d^2x^i}{dt^2} =&-\frac{1}{N}\left[\frac{1}{\sqrt{{\bar g}}}\,%
    {\bar g}_{j\ell}{\bar g}_{km}{\bar\pi}^{\ell m} \frac{dx^j}{dt}\frac{dx^k}{dt} 
    -  2\frac{dx^j}{dt} \partial_jN - \frac{\partial N}{\partial t}\right]%
    \frac{dx^i}{dt}
\qquad\hbox{at $p$}
\label{d5}
\end{align}
Accelerations are parallel to velocities, so  freely falling particles move in 
straight lines, but their proper times do not agree with the preferred 
time-slicing.  To make this quantitative, recall first that if $F$ is a function 
of space and time, its derivative along a geodesic is
\begin{align}
\frac{dF}{dt} 
    = \frac{\partial F}{\partial t} + \frac{dx^k}{dt}\partial_k F
\label{d6}
\end{align}
On the other hand, using (\ref{d2})--(\ref{d4}) and keeping in mind that 
$N^i=0$ at $p$, we have
\begin{multline}
  -\frac{1}{N}\left[\frac{1}{\sqrt{{\bar g}}}\,%
    {\bar g}_{j\ell}{\bar g}_{km}{\bar\pi}^{\ell m} \frac{dx^j}{dt}\frac{dx^k}{dt} 
    -  2\frac{dx^j}{dt} \partial_jN - \frac{\partial N}{\partial t}\right]\\
     = -\frac{1}{N}\biggl[\frac{1}{N}{\bar g}_{j\ell}{\bar\nabla}_kN^\ell\frac{dx^j}{dt}\frac{dx^k}{dt} 
     - \frac{dx^j}{dt}\partial_jN - \frac{dN}{dt}\biggr]  \label{d7} \\
    =-\frac{1}{N}\biggl[\frac{1}{N}\frac{dx^j}{dt}{\bar g}_{j\ell}\left(\frac{dN^\ell}{dt}
    - \frac{\partial N^\ell}{\partial t}\right)  - \frac{dx^j}{dt}\partial_jN 
    - \frac{dN}{dt}\biggr] \\
    = \frac{1}{N}\biggl[ \frac{dN}{dt} 
    - \frac{1}{N}{\bar g}_{j\ell}\frac{dx^j}{dt}\frac{dN^\ell}{dt}\biggr] 
    = \frac{1}{\lambda}\frac{d\lambda}{dt}  \qquad\hbox{at $p$}
\end{multline}
with
\begin{align}
\lambda = N - \frac{N_k}{N}\frac{dx^k}{dt}
\label{d8}
\end{align}
Hence we can write (\ref{d5}) as
\begin{align}
\frac{d^2x^i}{d\tau^2} = 0 \quad \hbox{with}\quad 
    d\tau = \lambda dt = Ndt - \frac{1}{N} N_kdx^k
\label{d9}
\end{align}

We thus recover ordinary inertial motion, provided that clocks are synchronized
to read the time $\tau$ rather than the preferred time given by the maximal
slicing of shape dynamics.   In fact, $\tau$ may be recognized as the standard 
``time'' orthogonal to the preferred spatial foliation of shape dynamics, in the
sense that the full spacetime interval is 
\begin{align}
ds^2 = d\tau^2 - {\bar g}_{ij}dx^idx^j
\label{d10}
\end{align}

So far, though, the straight line motion of (\ref{d9}) is compatible with 
either Lorentz invariance or Galilean invariance.  To distinguish the two,
we must also look at the existence and structure of light cones.  Let us 
define
\begin{align}
\sigma = \left(\frac{ds}{dt}\right)^2 = \left(\frac{d\tau}{dt}\right)^2 
     -  {\bar g}_{ij}\frac{dx^i}{dt}\frac{dx^j}{dt}  
    = \lambda^2 -  {\bar g}_{ij}\frac{dx^i}{dt}\frac{dx^j}{dt}
\label{d11}
\end{align}
where by (\ref{d10}), the sign of $\sigma$ distinguishes timelike, null, 
and spacelike geodesics in general relativity, with $\sigma=0$ for null 
geodesics.   

Fixing $\sigma$ at the point $p$ and differentiating (\ref{d11}), we have 
\begin{align}
\frac{d\sigma }{dt}  = 2\lambda\frac{d\lambda}{dt} 
    - 2 {\bar g}_{ij}\frac{d^2x^i}{dt^2}\frac{dx^j}{dt}
    - \left(\partial_t{\bar g}_{ij} + \partial_k{\bar g}_{ij}\frac{dx^k}{dt}\right)
    \frac{dx^i}{dt}\frac{dx^j}{dt}   \qquad\hbox{at $p$}
\label{d12}
\end{align}
The last term in this expression vanishes by virtue of (\ref{d3}) and (\ref{d4}),
and by using (\ref{d5})--(\ref{d7}) we obtain
\begin{align}
\frac{d\sigma }{dt}  = 2\lambda\frac{d\lambda}{dt} - 2{\bar g}_{ij}\frac{dx^j}{dt}
    \left(\frac{1}{\lambda}\frac{d\lambda}{dt}\frac{dx^i}{dt}\right)
   = \left(\frac{2}{\lambda}\frac{d\lambda}{dt}\right)\sigma
    \qquad\hbox{at $p$}
\label{d13}
\end{align}

It is evident from (\ref{d13}) that if $\sigma$ vanishes at $p$, it remains 
zero over the whole trajectory.  The condition $\sigma =0$ thus 
characterizes the entire path, and allows us to label it as a null geodesic.  
Since the metric ${\bar g}_{ij}$ is positive definite, the condition $\sigma=0$ 
determines a celestial sphere of vectors at $p$; given any initial 
direction, it defines a trajectory in that direction at the ``speed of light.''
This is precisely the Hamiltonian picture of a null cone at $p$.  Moreover,
it is clear that geodesics with $\sigma\ne0$ at $p$ cannot cross this 
cone---if $\sigma$ vanishes anywhere on a geodesic, it vanishes 
everywhere---so we obtain the usual picture of timelike, null, and spacelike 
regions.

Thus, once again, shape dynamics allows us to recover a form of Lorentz 
invariance.  But once again, the result is quite a bit less direct than it would 
have been in general relativity.   Note also that this recovery of local Lorentz 
invariance is contingent on $N$ being nonzero at $p$; if $N=0$, then 
shape dynamics and general relativity lose their equivalence, and Lorentz 
invariance is no longer expected.

\section{Conclusions}

Under most circumstances, shape dynamics is classically equivalent to 
general relativity: solutions of equations of motion of shape dynamics
in a particular conformal gauge are locally equivalent to solutions of the
Einstein field equations in a particular coordinate system.  It is therefore 
not surprising that Lorentz invariance can be found hidden inside
shape dynamics.  But while general relativity, with its manifest spacetime 
structure, exhibits Lorentz invariance simply and explicitly, the symmetry
must be teased out of shape dynamics.  It is, in one sense of the term,
emergent.

For weak fields and laboratory applications, this presumably makes standard
general relativity more convenient.  In other contexts, though, shape dynamics
may have an advantage.  For a spatially compact universe, the preferred time 
of shape dynamics is essentially the local Hubble expansion rate (averaged
over direction), suggesting that cosmological applications may simplify.  
In particular, gauge-invariant observables are much more easily identifiable 
in shape dynamics, and have a simple interpretation in terms of the spin 
decomposition of the extrinsic curvature and the metric fluctuations.  At first 
order in perturbation theory about an FLRW solution, for instance, a spin-zero 
degree of freedom can be constructed trivially, and corresponds precisely 
to the Mukhanov variable in constant mean curvature gauge \cite{cosmo}.

There is clearly more to be understood about the asymptotic symmetries
of shape dynamics as well.  Much of our intuition about asymptotic
symmetries comes from general relativity, and that intuition underlies 
the treatment of asymptotically flat spacetimes we have discussed here.  
It seems very difficult to enlarge the asymptotic symmetry group to include 
conformal transformations, but it may be possible to find different boundary 
conditions that allow asymptotic conformal transformations in place of 
boosts.  Geometries with anisotropic asymptotic scaling behavior have 
recently come under scrutiny \cite{Horava, Ross}; these might have 
interesting variants in shape dynamics.  

The most interesting questions, however, comes from the places in which 
shape dynamics and general relativity are \emph{not} equivalent.  In 
particular, when the lapse function (\ref{a6}) goes to zero---at the horizon 
of a stationary black hole, for instance---the correspondence with general 
relativity breaks down, and solutions can be physically different 
\cite{SDbh,SDbh2}.  Similarly, certain singularities in general relativity 
may be nonsingular in shape dynamics.  In these situations, the 
``emergent'' Lorentz invariance of shape dynamics is expected to 
disappear, with consequences that are only starting to be explored.

\vspace{1.5ex}
\begin{flushleft}
\large\bf Acknowledgments
\end{flushleft}

We would like to thank Gabe Herczeg and Vasudev Shyam for helpful conversations.
This work was supported in part by Department of Energy grant
DE-FG02-91ER40674.

\end{document}